# Robot mirroring: A framework for self-tracking feedback through empathy with an artificial agent representing the self

Monica Perusquía-Hernández[1], David Antonio Gómez Jáuregui[2], Marisabel Cuberos-Balda[3], Diego Paez-Granados[4]

[1]NTT Communication Science Laboratories, [2] Univ. Bordeaux, ESTIA, [3]Independent Researcher, [4]Swiss Federal School of Technology in Lausanne - EPFL
`perusquia@ieee.org, d.gomez@estia.fr,`
`marisabelcb@gmail.com, dfpg@ieee.org`

**Abstract.** Current technologies have enabled us to track and quantify our physical state and behavior. Self-tracking aims to achieve increased awareness to decrease undesired behaviors and lead to a healthier lifestyle. However, inappropriately communicated self-tracking results might cause the opposite effect. In this work, we propose a subtle self-tracking feedback by mirroring the self's state into an artificial agent. By eliciting empathy towards the artificial agent and fostering helping behaviors, users would help themselves as well. Finally, we reflected on the implications of this design framework, and the methodology to design and implement it. A series of interviews to expert designers pointed out to the importance of having multidisciplinary teams working in parallel. Moreover, an agile methodology with a sprint zero for the initial design, and shifted user research, design, and implementation sprints was proposed. Similar systems with data flow and hardware dependencies would also benefit from the proposed agile design process.

**Keywords:** Self-quantification, robotics design, stress monitoring

## 1 Introduction

Current technology has allowed us to precisely measure several phenomena, including the self. Self-tracking includes physiological states such as Heart Rate (HR), Respiration Rate (RR), Electro-Dermal Activity (EDA), among others. Bodily states are often measured with the goal of monitoring our health and timely detect anomalies to treat them on time. Models of Personal Informatics (PI) work under the assumption that self-tracking leads to self-insight, and eventually to self-improvement or a positive behavior change (Self-Improvement Hypothesis) [1]. Moreover, behavior change models such as the Trans-Theoretical Model of Behavior Change (TTM) [2] imply that increased awareness includes two elements: awareness of an unknown fact, and awareness of reasons why behavior change is advantageous. In other words, both awareness of the problem and the motivations to solve it are necessary elements for behavior change. Therefore, for PI systems to induce behavioral changes, the motivations given for behavior change should outweigh reasons favoring staying the same. Additionally, for long lasting behavior change, the perception of self-efficacy in achieving the desired goals is necessary, as well as temporally near feedback on the relevant behavior [3].

Recent sensing technology and mobile applications have enabled us to monitor and log our behavior to immediately obtain feedback on the monitored state. For example, HR could be directly displayed to the screen, and the user could interpret the given number contextually according to the activity he is realizing. This would help users to identify co-occurring events that cause HR variations. Other cues that can add to the interpretability of HR can also be tracked simultaneously. For example, HR and respiration rates are known to depend on each other, and respiration exercises are recommended to control one's heart [4, 5]. Even though technology has enabled us to increase the number of tracked features, its interpretation might be complex and might be misinterpreted by a naïve user. Moreover, for each type of information, a modality or a channel is required to provide feedback. However, it is difficult for a human to handle as many channels, and more technically complex to implement them. Therefore, appropriate data visualization comes into play to deliver relevant, behavior changing information. Nevertheless, the challenge of delivering feedback concisely and avoiding misinterpretation remains. For example, people might only believe information that confirms or suits their beliefs [6]. These confirmation biases may skew the interpretation of raw data feedback. On the other hand, machines excel at the task of processing high amounts of information. Recent machine learning techniques can be used to automatically match raw information to a human-understandable concept. We usually do not think about how many beats per second our heart pumps. Instead, we think about how that makes us feel, and we name those feelings. Nevertheless, data interpretation by artificial intelligence algorithms might not be perfect and situations of distrust or over-trust might arise [6]. Therefore, high levels of automation in data interpretation and user feedback should be considered carefully. For example, incorrectly detecting that someone is angry and stating so might lead to two outcomes. Either the user's trust on the feedback decreases because it was incorrect, or a self-fulfilling prophecy is created when the users get angry when being told they are angry. Usually, the degree in which a machine learning system can tolerate false alarms can be controlled a priori, and it strongly depends on the system application. It might be better to falsely diagnose a disease than to miss the symptoms and not to provide proper treatment to a sick person. Hence, the feedback provided to a user who wants to lose weight because of aesthetic reasons should be different from the feedback provided to a diabetic patient at risk of complications. In non-medical conditions, users might benefit from more lenient feedback thresholds.

Despite most theories for self-improvement through self-tracking assume awareness is necessary, we argue that, sometimes, a lack of awareness might be more beneficial. Overflowing the user with information that is not 100% accurate might boost the aforementioned unintended effects. Therefore, we propose quantification feedback delivered subtly, without creating awareness. The users does not necessarily need to see each measure, but rather a reflection of their health in an embodied character. This is similar to looking at oneself in a mirror, and it would allow users to better interpret their state. Moreover, by provoking empathy with the robot, we also foster self-compassion. Self-compassion might also trigger healthier behaviors [8, 9, 10, 11]. As human beings, we are largely influenced and inspired by others. A naïve desire to take care of, and give a good example to our children might motivate us to live a healthier lifestyle. Therefore, we hypothesize that behavior change can be fostered without necessarily passing through the awareness bridge stated by the Self-Improvement Hypothesis. In the following sections we provide evidence from the literature to support our robot mirroring approach to self-tracking feedback. Furthermore, we used expert interviews as information source to outline an appropriate design methodology to successfully implement a design using the robot mirroring concept.

## 2 Embodiment of the user state into an artificial agent (robot) to foster empathy towards it, and indirectly, self-compassion

Empathy has been defined as three dimensions of understanding others: "perspective taking, which is the tendency to take on the point of view of others in interpersonal situations; empathic concern, which is the tendency to experience feelings of care and concern towards others; and personal distress, which is the tendency to react with discomfort to the emotional experience of others." [12] Understanding of others emotions is important to foster compassionate behaviors to support each other [13]. Caring and supporting others often leads to a sense of satisfaction and fulfillment that is beneficial for both self- and others- wellbeing [14]. When speaking only about own emotions, the terms used are *mindfulness* for self-understanding, and *self-compassion* for understanding oneself and treating oneself with kindness, care, and concern in the face of negative life events [9]. Self-compassion has been related to promoting successful self-regulation of health-related behaviors [8]. Moreover, framing medical problems and their treatment in ways that foster self-compassion may enhance people's ability to manage their health-related behavior and deal with medical problems [10]. Thus, being self-compassionate often leads to a healthier life-style and may affect our self-efficacy sense to achieve our behavior change goals.

In our proposal, we aim to improve human well-being by encouraging well-being directed actions towards an embodied agent, or robot. This implies that the robot design should foster clear communication and empathetic understanding with the user. Moreover, when the robot mirrors the health state of the user, self-compassion would be indirectly encouraged.

Empathetic behaviors towards robots has indeed been explored in previous research. In a longitudinal study, [15] found out that children took care of a robot that suddenly ran out of battery as if the robot were sleeping. This behavior is more salient for robots that are somehow perceived as weaker than the user. Because of this, feeble robots can be used as supporting tools in education settings [16]. This concept is often referred as care-receiving robots. A more intuitive example is that people often overestimate their ability to cope with difficult situations [17, 18]. Thus, they do not take preventive strategies. However, when there is a person with whom we empathize, and this person is perceived as unable to cope or is akin to us (i.e., our child), we would tend to take caring strategies to help them adopt healthier behaviors [19, 20, 21]. In this line of thought, self-tracking can be used to identify opportunities for wellbeing improvement. Feedback can be provided subtly to avoid unintended effects of self-tracking such as excessive awareness and over-trust. Finally, by fostering empathy and helping behaviors towards an artificial agent mirroring the self, we aim to indirectly foster self-compassion and self-helping behaviors that improve the user's wellbeing.

## 3 Paradigm shifting for embodied agent design

A robot is an embodied agent whose interactions to cope with the environment depend largely on its body sensing capabilities. Because of this, software, behavior, and interactions with the environment and other agents strongly depend on its hardware design. A large amount of research in the Human-Robot Interaction community uses out-of-the-shelf robots that were designed with a general purpose. The number of robots designed completely for a specific purpose is more limited. For example, [22] proposed to design robots considering expected movements to interact with people even before the hardware design. This is of utmost importance when designing robots with a specific application. Such robots would probably require specialized embodied characteristics. Nevertheless, full design of specialized robot hardware poses a challenge. Whereas current design and implementation cycles can be adapted into an agile methodology, robot design has been prominently made following a waterfall methodology.

In this section we describe related works and the design methodology used. Afterwards, we describe a series of semi-structured interviews aimed to unveil the design process followed by experts in UI/UX design, computer science and mechatronics. Based on this information, we propose a tentative design process for robots with requirements similar to the one we are proposing.

### 3.1 Literature review

In social agents' research, several authors have proposed embodied agents capable of producing empathy and self-compassion responses with the aim of fostering changes in users' behavior. To show caring and empathetic behavior, an agent must be attuned to the affective state of the user [23]. Moreover, artificial agents can be categorized in both types: virtual agents and robots [24]. One of the first works regarding empathic virtual agents was proposed by [25]. They proposed an interactive pedagogical drama in which a user interacts with virtual characters in a believable story that the user empathizes with. The goal of this interaction is to help mothers of very sick children to manage her own stress and to improve problem-solving skills. Similarly, [26] proposed an application of virtual drama to help children address bullying situations. The virtual drama used empathetic responses to virtual characters to teach children different coping strategies and techniques for dealing with bullying in an individual and personal manner. In the same year, [27] proposed an Embodied Conversational Agent (ECA) that tried to emulate, with a Wizard of Oz, the behavior of a human therapist to promote interest in healthy eating to the users. More recently, [28] proposed a virtual reality game in which users controlled a superhero character to fulfill altruistic tasks. Results indicated that the videogame led user to greater prosocial behavior in the real world. One year later, [29] proposed a virtual human that acts as an interviewer designed to address problems of depression, anxiety, or post-traumatic stress disorder in patients. Similarly, [30] proposed a virtual agent with back story to train medical students overcome challenges associated with interpersonal communication skills. However, one problem of virtual agents is the limited physical interaction resulted from the use of a computer screen and predefined dialogues to inform the virtual agent the user's emotional state [24]. The interaction with robots tends to be more natural by allowing to adapt the robot's behavior to the user's emotional state during the interaction. Significant improvements in emotion recognition using different modalities have allowed robots to understand the user state, and thus, to produce empathetic responses [23]. In 2005, [31] proposed an empathetic anthropomorphic robot (torso) that shows happiness, fear and neutral facial expressions as responses to the affective speech signal of the users. The results suggest that users perceived the robot to react more adequately to emotional aspects of a situation ("situation fit"). Similarly, [32] proposed a robot with the form of a chimpanzee that mimics the facial expressions and head movements of users. The users considered the interaction more satisfactory than participants who interacted with a version of the robot without mimicking capabilities. Furthermore, [33] confirmed with an expressive robot head, that helping behaviors towards a robot can be increased by proactively adapting the robot's behavior to the mood of the user. More recently, [34] explored the role of empathy in long-term interaction between children and social robots. They developed a social robot (named iCat) that plays chess with children. The proposed social robot generated supportive behaviors according to the detected affective state of the children.

Although previous work in artificial agent design has actively encouraged empathy to deal with own distress or to promote social behaviors, none of them explored what would be the effect of mirroring the user's health state in the artificial agent to promote self-helping behaviors.

### 3.2 The recommended design and development process of new technologies according to experts

Designed for a general-purpose robot allows for a quick start to iterate a concept. Nevertheless, new behaviors are constrained by the hardware of the robot they are designed for. For our proposed concept, it would be nice to design a robot that can track users and provide them with feedback in a continuous manner. An alternative would be to embed the sensors in the robot, and in turn, making the robot a wearable. Available robots in the market do not fulfill these characteristics. Therefore, simultaneous UI/UX and technical design would be required. To identify critical steps necessary to design artificial agents from scratch in a multidisciplinary team, a semi-structured interview was conducted with six professionals (mean: 33.3 yo, SD: 3.2, 2 female). Their expertise was in robotics engineering, UI/UX design, Human-Robot Interaction (HRI), and software engineering (average years of experience: 8.3, SD: 5.0). The interview had two sections. One about general design experiences and their design workflow when creating robots. The second one asked directly about the design of a robot embodying the user state. The responses were transcribed and analyzed by creating Affinity Diagrams.

All of the professionals who participated in the interview used iterative design and development processes, regardless of their profession. The duration of each cycle is predefined beforehand. This time boxing is critical and supports prioritization (P4, 10 years of experience in both UI/UX design and software development). Although the design process is iterative and agile, the first iteration is still in waterfall (sprint zero) because the design and its respective user research should go one sprint ahead of the technical development (P1, 2 years of experience in Human-Robot Interaction design; P2, 8 years of experience in UI/UX design; P4). In this first design round the conceptual design should be discussed among designers, social scientists, and engineers. The Minimum Viable Product (MVP) is a team decision (P4) determined in sprint zero (P2-4) considering what can bring value to the users (P4).

The design of a robot requires a multidisciplinary team (P1; P2; P3, 18 years of experience in software engineering and HRI; P5, 5 years of experience in robotics; P6, 7 years of experience in software research for HRI). The interviewees suggested at least four roles: (1) Industrial and interaction designers designing the human interface; (2) Mechatronics engineers dealing with the electrical and mechanical part; (3) Software engineers able to deal with data analysis and behavior programming; (4) Social scientists such as psychologists or neuroscientists to deal with user research, and to provide insights from the human mind. To overcome misunderstandings across disciplines, concepts should be communicated with clarity and simplicity (P2-3). Sketches are useful in this task (P1, P2, P4). Clay is also useful to envision the dimensions of the robot (P1).

Usually, designers focus more on solving user needs through user research without getting involved in technical complexities (P2). Some only focus on the appearance of the robot (P1). Although designers usually value and deal with abstract concepts, these are not comfortable for engineers (P2). Engineers require specific and detailed requirements. Thus, the final design in an iteration should be defined and translated to functional requirements with specific inputs and outputs (P3) (Fig. 1). It is important for engineers to understand the big picture and get the requirements, but they should not be told how to make the implementation. For both designers and engineers, it is easier to accept feedback on their work from someone knowledgeable on their fields than from someone with different background. Therefore, project owners with general knowledge about both are required as a bridge between teams and to distyle the concept into small tasks (P1-4).

There are many dependencies in the development of an artificial agent. Therefore, both design and technological development should be done in close communication (P4). The physical robot design depends on the expected behavior and robot movements. Afterwards, the software, serving as the mind of the robot, can be implemented (P1). To sort out these dependencies, adaptations and interactions with the environment should be defined first (P3), followed by sensing modules with well-defined inputs and outputs (P6, P3). Moreover, modularity with clear functional requirements might enable the development team to parallelize tasks (P4-6). Fig. 2 shows this process.

Finally, it is very challenging to build low-fidelity prototypes for user testing during new technology design. *"I often wonder what is faster, coding or doing a mockup ... Especially if data is involved, I often go for the coded prototype. It is difficult to make a fixed scenario for a complex data search and visualization"* (P4). In those cases, both user and data flows must be considered. Data and its storage are part of the design process. Therefore, high fidelity, functional prototypes that require real data should be tested iteratively (P4).

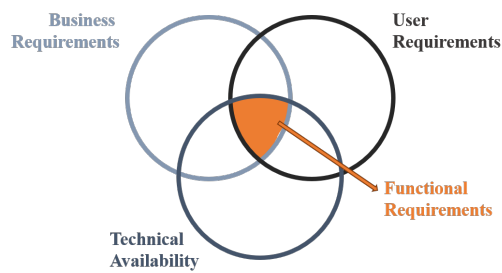

**Fig. 1.** Functional requirements are key to maintain clarity and good communication between design and engineering teams. They are a compromise between business requirements, user requirements and the feasibility to implement the technologies required to realize a concept.

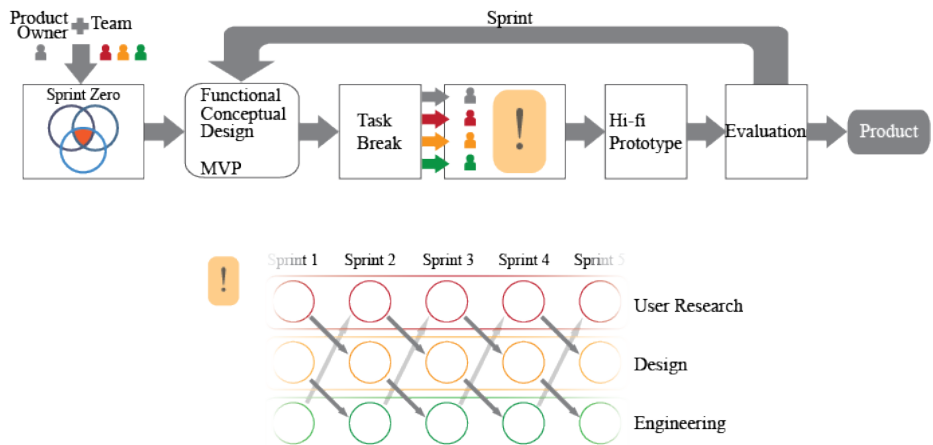

Fig. **2**. Suggested design-implementation-evaluation process for robot platforms. An agile methodology allows to accelerate the process and to evaluate the product iteratively. In a sprint zero, the Minimum Viable Product (MVP) is defined by all team members considering the initial requirement. This concept is broken into the initial functional requirements that the engineering team use to start the implementation. In the meanwhile, User research teams work on the evaluation of previous prototypes, and UI/UX designers work on improving the concept for the next iteration.

## 3.3 Fast Technical Prototyping as a Key in the Design Process

The experts' opinions regarding the importance of iterative prototyping lead to an unavoidable bottle neck: the first functional prototype with enough capabilities to be used in user-studies. In the case of a robot, this prototype would be built by a mechatronic designer. Mechatronic design is a cross-discipline work that considers, mechanical, electronic, control and software disciplines to achieve an intended product [35]. As in other design cycles, conceptual prototypes and virtual prototypes (model-based), are necessary before building an initial product [36]. However, as part of the global design process of a Human-Robot Interaction, we would recommend going ahead with conceptual prototypes in conjunction or with support of the UI/UX team. The modeling phase can also be skipped to focus on integrating a holistic system that uses current of-the-shelf software architectures, control methods, and electronics. All of them implemented in the simplest possible version of a functional mechanical device. Moreover, current approaches for fast prototyping rely on 3D printing techniques [37] for accelerating the mechanical component design. This allows great flexibility and expands the level of complexity that can be achieved even on the initial prototype. However, the level of proficiency of the mechanical designer in this technique should be considered, as it requires a significant amount of virtualization of parts and overall components. Thus, other alternatives should not be immediately taken off the table, such as, hand-made models with clay and wood; assembled models with basic structures such as aluminum frames; or of-the-shelf component assembly. The main criteria for choosing the level of complexity of the initial functional model should be based on the requirement of the UX/UI team for their future iteration process with users. This translates in creating a specific requirement traceability matrix (RTM) [38], that should ignore some of the final requirements (costs, sizes, weights, optimal structures, novelty), and introduce fast-prototyping as key requirement with the highest weighting.

## 4 Case-study: Stress quantification

As previously mentioned, most of behavior change theoretical models assume that awareness of a problem is the first step to solve it [39]. The same has been assumed for stress. If we are aware of stress, we can reduce it. This assumes that most people are not aware of how much stress they have and what causes the stress. By tracking it, people can become more aware of its causes [1]. Thus, more awareness might lead to reduced stress and a happier, healthier life. Nevertheless, it can be that some people with neurotic traits get more stressed than others, especially if they are over self-conscious [40]. As previously mentioned, self-tracking might have unintended effects. People might get stressed when they are explicitly told they are stressed. On the other hand, if the feedback is more abstract, people might interpret its meaning to maintain their existing beliefs [7], and the belief that stress is bad might be more harmful than the stress itself [42]. Therefore, re-appraisals on the meaning of body stress and its relationship to stressors is believed to be a successful strategy when dealing with it [43]. There is also some evidence that stress prompts social responses that create resilience against it [44]. These responses include increased physical contact, empathy, compassion, and caring. When projecting such social needs into an artificial intelligence agent such as a pet robot that mirrors the users themselves, we expect to trigger a beneficial cycle that might allow users to reduce their stress levels. This can be achieved without necessarily becoming aware of the quantification, and the artificial intelligence algorithms designed to measure stress.

### 4.1   A wearable pet robot for stress management

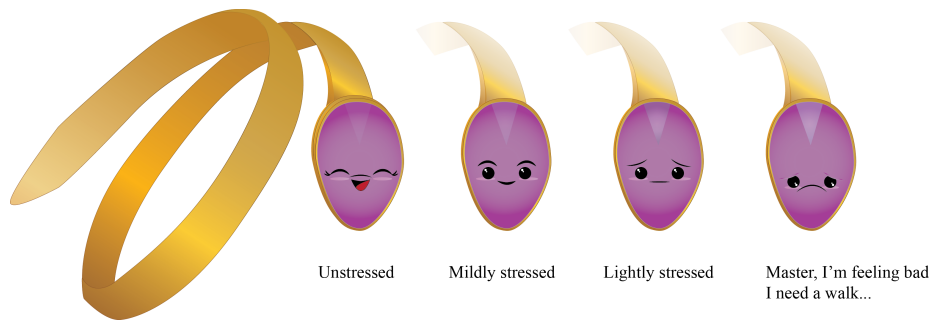

Unstressed    Mildly stressed    Lightly stressed    Master, I'm feeling bad
                                                      I need a walk...

Fig. **3**.  Pet robot feedback concept states. The robot should measure the user's stress level and provide subtle feedback about the user state by mirroring it. Our hypothesis is that by having empathy towards the pet robot, the user would develop self-helping behaviors.

Given these constraints, we propose an instance of empathetic self-tracking robot: a pet robot that mirrors the user's health state as a mean to provide feedback (Fig. 3). The following design elements should be considered and detailed using the methodology proposed in the previous section.

1. The robot should measure the user's state through several sensors arranged in a wearable for comfortable.
2. The data obtained should be processed to roughly estimate the user's stress state.
3. The robot should be able to mirror several levels of the user's health state.
4. The robot should appear as weaker than the user, to foster care-receiving treats.
5. The robot should trigger empathy from the user and exhibit empathetic behaviors towards the user.
6. The robot should track successes and failures to encourage helping behaviors, and increasingly adapt to improve these.

## 5   Conclusion and future work

We challenge the assumption that awareness is required to induce behavior change. We proposed a robot mirroring approach to create empathy and behavior change by fostering helping behaviors towards the artificial agent, and indirectly, self-compassion. Moreover, we inquired several experts to outline an agile methodology for this and similar robotic platforms to successfully be designed, implemented, and evaluated. Future work should generate more specific user requirements, and concretize the design, implementation, and evaluation of the proposed robot.

## References

1. Kersten - van Dijk, E. T. (2018). Quantified stress : toward data-driven stress awareness. Eindhoven: Technische Universiteit Eindhoven.
2. Prochaska, J. O., & Velicer, W. F. (1997). The Transtheoretical Model of Health Behavior Change. American Journal of Health Promotion, 12(1), 38–48. https://doi.org/10.4278/0890-1171-12.1.38.
3. Bandura, A. (1978). Self-efficacy: Toward a unifying theory of behavioral change. Advances in Behaviour Research and Therapy, 1(4), 139–161
4. Gilbert, C. (2003). Clinical Applications of Breathing Regulation. *Behavior Modification*, *27*(5), 692–709.
5. Yu, B., Feijs, L., Funk, M., & Hu, J. (2015). Breathe with Touch: A Tactile Interface for Breathing Assistance System (pp. 45–52). Springer, Cham.
6. Klayman, J. (1995). Varieties of Confirmation Bias. Psychology of Learning and Motivation, 32, 385–418.
7. van Dijk, E. T., Beute, F., Westerink, J. H. D ., & Ijsselsteijn, W. A. (2015). Unintended effects of self-tracking. CHI'15, April 18–April 23, 2015, Seoul, South-Korea. Workshop on 'Beyond Personal Informatics: Designing for Experiences of Data,' 5.

8. Raab, K. (2014). Mindfulness, Self-Compassion, and Empathy Among Health Care Professionals: A Review of the Literature. Journal of Health Care Chaplaincy, 20(3), 95–108
9. Neff, K. (2003). Self-Compassion: An Alternative Conceptualization of a Healthy Attitude Toward Oneself. Self and Identity, 2(2), 85–101.
10. Terry, M. L., & Leary, M. R. (2011). Self-compassion, self-regulation, and health. Self and Identity, 10(3), 352–362.
11. Birnie, K., Speca, M., & Carlson, L. E. (2010). Exploring self-compassion and empathy in the context of mindfulness-based stress reduction (MBSR). Stress and Health, 26(5), 359–371.
12. Kingsbury, E. (2009). The relationship between empathy and mindfulness: Understanding the role of self-compassion. - PsycNET.Dissertation Abstracts International: Section B: The Sciences and Engineering,70(5–B), 3175.
13. Paciello, M., Fida, R., Cerniglia, L., Tramontano, C., & Cole, E. (2013). High cost helping scenario: The role of empathy, prosocial reasoning and moral disengagement on helping behavior. Personality and Individual Differences, 55(1), 3–7.
14. Hojat, M. (2007). Empathy in patient care : antecedents, development, measurement, and outcomes. Springer.
15. Tanaka, F., Cicourel, A., & Movellan, J. R. (2007). Socialization between toddlers and robots at an early childhood education center.Proceedings of the National Academy of Sciences of the United States of America,104(46), 17954–17958.
16. Tanaka, F., & Kimura, T. (n.d.).Care-receiving Robot as a Tool of Teachers in Child Education. Retrieved from http://fumihide-tanaka.org/old/paper/Tanaka_IS-10.pdf
17. Weinstein, N. D. (1982). Unrealistic optimism about susceptibility to health problems. Journal of Behavioral Medicine, 5(4), 441–460.
18. Weinstein, N. D. (1980). Unrealistic optimism about future life events. Journal of Personality and Social Psychology, 39(5), 806–820.
19. Dovidio, J. F., & Penner, L. A. (2008). Helping and Altruism. In G. J. O. Fletcher & M. S. Clark (Eds.), Interpersonal Processes (pp. 162–195). John Wiley & Sons.
20. Konečni, V. J., & Ebbesen, E. B. (1975). Effects of the Presence of Children on Adults' Helping Behavior and Compliance: Two Field Studies. The Journal of Social Psychology, 97(2), 181–193. https://doi.org/10.1080/00224545.1975.9923338
21. Dovidio, J. F. (1984). Helping Behavior and Altruism: An Empirical and Conceptual Overview. Advances in Experimental Social Psychology, 17, 361–427.
22. Hoffman, G., Zuckerman, O., Hirschberger, G., Luria, M., & Shani-Sherman, T. (2015). Design and Evaluation of a Peripheral Robotic Conversation Companion. In ACM/IEEE International Conference on Human-Robot Interaction (HRI). IEEE.
23. Picard, R. (1997). Affective Computing. Cambridge, MA: MIT Press.
24. Paiva, A. and Leite, I. and Boukricha, H. and Wachsmuth, I. (2017). Empathy in Virtual Agents and Robots: A Survey. ACM Trans. Interact. Intell. Syst., 7(3), 11:1--11:40.
25. Stacy C. Marsella, W. Lewis Johnson, and Catherine LaBore. 2000. Interactive pedagogical drama. In Proceedings of the Fourth International Conference on Autonomous Agents. ACM, 301–308.
26. Aylett R.S., Louchart S., Dias J., Paiva A., Vala M. (2005) FearNot! – An Experiment in Emergent Narrative. In: Panayiotopoulos T., Gratch J., Aylett R., Ballin D., Olivier P., Rist T. (eds) Intelligent Virtual Agents. IVA 2005. Lecture Notes in Computer Science, vol 3661. Springer, Berlin, Heidelberg
27. Fiorella de Rosis, Addolorata Cavalluzzi, Irene Mazzotta, and Nicole Noviielli. 2005. Can embodied conversational agents induce empathy in users? Virtual Social Agents (2005), 65.
28. Robin S. Rosenberg, Shawnee L. Baughman, and Jeremy N. Bailenson. 2013. Virtual superheroes: Using superpowers in virtual reality to encourage prosocial behavior. PloS One 8, 1 (2013).
29. David DeVault, Ron Artstein, Grace Benn, Teresa Dey, Ed Fast, Alesia Gainer, Kallirroi Georgila, Jon Gratch, Arno Hartholt, Margaux Lhommet, and others. 2014. SimSensei kiosk: A virtual human interviewer for healthcare decision support. In Proceedings of the 2014 International Conference on Autonomous Agents and Multi-agent Systems. International Foundation for Autonomous Agents and Multiagent Systems, 1061–1068.
30. Andrew Cordar, Michael Borish, Adriana Foster, and Benjamin Lok. 2014. Building virtual humans with back stories: Training interpersonal communication skills in medical students. In Intelligent Virtual Agents. Springer, 144–153.
31. Frank Hegel, Torsten Spexard, BrittaWrede, Gernot Horstmann, and Thurid Vogt. 2006. Playing a different imitation game: Interaction with an empathic android robot. In Proceedings of the 2006 IEEE-RAS International Conference on Humanoid Robots (Humanoids06). 56–61.
32. Laurel D. Riek, Philip C. Paul, and Peter Robinson. 2010. When my robot smiles at me: Enabling human-robot rapport via real-time head gesture mimicry. Journal on Multimodal User Interfaces 3, 1–2 (2010), 99–108.
33. Barbara Gonsior, Stefan Sosnowski, Malte Buß, DirkWollherr, and K. Kuhnlenz. 2012. An emotional adaption approach to increase helpfulness towards a robot. In Intelligent Robots and Systems (IROS), 2012 IEEE/RSJ International Conference on. IEEE, 2429–2436.
34. Iolanda Leite, Ginevra Castellano, André Pereira, Carlos Martinho, and Ana Paiva. 2014. Empathic robots for long-term interaction: Evaluating social presence, engagement and perceived support in children. International Journal of Social Robotics (2014), 1–13.
35. Jovanovic, V., Michaeli, J. G., Popescu, O., Moustafa, M. R., Tomovic, M., Verma, A. K., & Lin, C. Y. (2014). Implementing mechatronics design methodology in mechanical engineering technology senior design projects at Old dominion university. ASEE Annual Conference and Exposition, Conference Proceedings.
36. Gausemeier, J., Dumitrescu, R., Kahl, S., & Nordsiek, D. (2011). Integrative development of product and production system for mechatronic products. Robotics and Computer-Integrated Manufacturing, 27(4), 772–778.
37. Jasveer, S., & Jianbin, X. (2018). Comparison of Different Types of 3D Printing Technologies. International Journal of Scientific and Research Publications (IJSRP), 8(4), 1–9.
38. Gotel O. et al. (2012) Traceability Fundamentals. In: Cleland-Huang J., Gotel O., Zisman A. (eds) Software and Systems Traceability. Springer, London.
39. Prochaska, J. O., & Velicer, W. F. (1997). The Transtheoretical Model of Health Behavior Change.American Journal of Health Promotion,12(1), 38–48.
40. Schneider, T. R. (2004). The role of neuroticism on psychological and physiological stress responses.Journal of Experimental Social Psychology,40(6), 795–804.
41. Osorio, L. C., Cohen, M., Escobar, S. E., Salkowski-Bartlett, A., & Compton, R. J. (2003). Selective attention to stressful distracters: effects of neuroticism and gender.Personality and Individual Differences,34(5), 831–844.
42. Keller, A., Litzelman, K., Wisk, L. E., Maddox, T., Cheng, E. R., Creswell, P. D., & Witt, W. P. (2012). Does the perception that stress affects health matter? The association with health and mortality.Health Psychology,31(5), 677–684.
43. Jamieson, J. P., Nock, M. K., & Mendes, W. B. (2012). Mind over matter: Reappraising arousal improves cardiovascular and cognitive responses to stress. Journal of Experimental Psychology: General, 141(3), 417-422.
44. Poulin, M. J., Brown, S. L., Dillard, A. J., & Smith, D. M. (2013). Giving to others and the association between stress and mortality.American Journal of Public Health,103(9), 1649–1655.